\documentclass[prl,aps,twocolumn,preprintnumbers,amsmath,amssymb,superscriptaddress,showpacs]{revtex4-1}
\usepackage[colorlinks,bookmarks=true,citecolor=blue,linkcolor=red,urlcolor=blue,citecolor=blue]{hyperref}

\usepackage{graphicx}
\usepackage{subfigure}
\usepackage{amssymb,amsmath}
\usepackage{color}
\usepackage{lineno}
\UseRawInputEncoding

\begin{document}
\switchlinenumbers
\title{Spin Excitation Spectra of Anisotropic Spin-$1/2$ Triangular Lattice Heisenberg Antiferromagnets}

\author{Runze Chi}\email{These authors contributed equally to this work}
\affiliation{Beijing National Laboratory for Condensed Matter Physics and Institute of Physics,
Chinese Academy of Sciences, Beijing 100190, China.}
\affiliation{School of Physical Sciences, University of Chinese Academy of Sciences, Beijing 100049, China.}

\author{Yang Liu}\email{These authors contributed equally to this work}
\affiliation{Beijing National Laboratory for Condensed Matter Physics and Institute of Physics,
Chinese Academy of Sciences, Beijing 100190, China.}
\affiliation{School of Physical Sciences, University of Chinese Academy of Sciences, Beijing 100049, China.}

\author{Yuan Wan}
\affiliation{Beijing National Laboratory for Condensed Matter Physics and Institute of Physics,
Chinese Academy of Sciences, Beijing 100190, China.}
\affiliation{Songshan Lake Materials Laboratory, Dongguan, Guangdong 523808, China.}

\author{Hai-Jun Liao}\email{navyphysics@iphy.ac.cn}
\affiliation{Beijing National Laboratory for Condensed Matter Physics and Institute of Physics,
Chinese Academy of Sciences, Beijing 100190, China.}
\affiliation{Songshan Lake Materials Laboratory, Dongguan, Guangdong 523808, China.}

\author{T. Xiang}\email{txiang@iphy.ac.cn}
\affiliation{Beijing National Laboratory for Condensed Matter Physics and Institute of Physics, Chinese Academy of Sciences, Beijing 100190, China.}
\affiliation{School of Physical Sciences, University of Chinese Academy of Sciences, Beijing 100049, China.}
\affiliation{Beijing Academy of Quantum Information Sciences, Beijing, 100190, China.}

\begin{abstract}
Investigation of dynamical excitations is difficult but crucial to the understanding of many exotic quantum phenomena discovered in quantum materials. This is particularly true for highly frustrated quantum antiferromagnets whose dynamical properties deviate strongly from theoretical predictions made based on the spin-wave or other approximations. Here we present a large-scale numerical calculation on the dynamical correlation functions of spin-$1/2$ triangular Heisenberg model using a state-of-the-art tensor network renormalization group method. The calculated results allow us to gain for the first time a comprehensive picture on the nature of spin excitation spectra in this highly frustrated quantum system. It provides a quantitative account for all the key features of the dynamical spectra  disclosed by inelastic neutron scattering measurements for $\rm Ba_3CoSb_2O_9$, revealing the importance of the interplay between low- and high-energy excitations and its renormalization effect to the low-energy magnon bands and high-energy continuums. We identify the longitudinal Higgs modes in the intermediate-energy scale and predict the energy and momentum dependence of spectral functions along the three principal axes that can be verified by polarized neutron scattering experiments. Furthermore, we find that the spin excitation spectra weakly depend on the anisotropic ratio of the antiferromagnetic interaction.
\end{abstract}

\maketitle

\begin{figure*}[htp]
\centering
\includegraphics[width=0.8\textwidth]{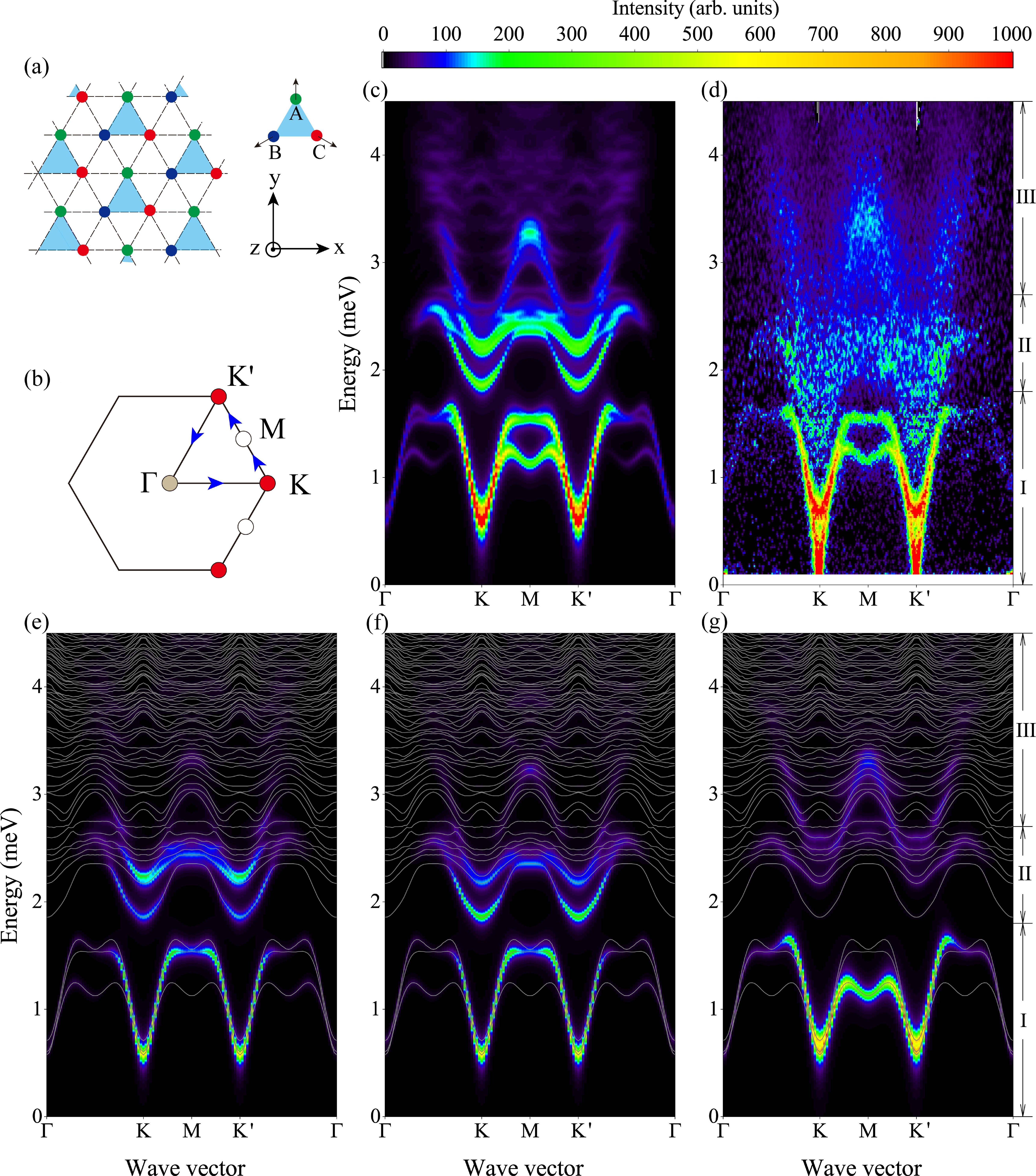}
\caption{
Comparison between the tensor-network results and the INS measurement data for the dynamical spin structure factors.
  (a) Triangular lattice and the $120^\circ$ N\'eel order in the ground state of the  Heisenberg model. The magnetization is ordered along the $y$-axis directon on sublattice $A$ in the triangular plane. (b) The first Brillouin zone and the momentum path (the arrowed lines) on which the spectral functions are evaluated. (c) The total spectral weight of the dynamical spectral function obtained with the tensor-network method for the easy-plane $XXZ$ model with $J=$1.67 meV and  $\Delta=$0.95. (d) The INS spectra of $\rm Ba_3CoSb_2O_9$ reproduced using the data published in Supplementary Materials in Ref. \cite{Macdougal2020}. (e)-(g) The spin structure factors along the three principal axes: (e) $S^{xx}(\boldsymbol{k},\omega)$, (f) $S^{yy}(\boldsymbol{k},\omega)$, and (g) $S^{zz}(\boldsymbol{k},\omega)$. Their sum gives the total spectral weight shown in (c). The gray curves show the energy dispersions of the excitation states. The spectra are divided into three stages according to their dispersions : (I) 0$<E<$1.8meV, (II) 1.8 meV $<E<$ 2.7 meV, and (III) $E>$2.7 meV.
\label{fig1}}
\end{figure*}

\textit{Introduction.---}Frustrated quantum magnetism has moved to the forefront of condensed matter physics research. Quite many exotic quantum phenomena driven by the interplay between quantum fluctuations and geometric frustrations, such as quantum spin liquid~\cite{Anderson1973, Balents2010, Zhou2017, Liao2017} and magnetic monopoles~\cite{Castelnovo2008}, have been discovered in these systems. 
The spin-1/2 triangular antiferromagnetic Heisenberg model is a prototypical frustrated magnetic system that has been intensively studied for more than four decades~\cite{Anderson1973,Fazekas1974,Huse1988,Capriotti1999,White2007}. While it is now commonly accepted that its ground state is noncollinear $120^{\circ}$ magnetic ordered [Fig.~\ref{fig1}(a)]~\cite{Huse1988,Capriotti1999,White2007}, the physical properties of its excitation states remain elusive.

The linear spin wave theory (LSWT) predicts that there are three magnon excitation modes in the triangular antiferromagnetic Heisenberg model. However, inelastic neutron scattering (INS) measurements on $\rm Ba_3CoSb_2O_9$~\cite{Ma2016, Ito2017, Macdougal2020}, which is an excellent realization of the spin-1/2 triangular Heisenberg model \cite{Susuki2013, Yamamoto2015, Shirata2012, Koutroulakis2015}, just observed two branches of magnon excitation modes. More surprisingly, these magnon excitation modes were found to be strongly renormalized around the $M$ point [Fig.~\ref{fig1}(d)] where the bands bend downward and one of them exhibits a rotonlike minimum. Moreover, two strong dispersive continuums of unknown origin are observed above the low-energy magnon bands~\cite{Ito2017,Macdougal2020}.

A number of theories have been proposed or invoked to explain the exotic magnetic spectra observed in $\rm Ba_3CoSb_2O_9$, based either on the multimagnon interactions~\cite{Zheng2006, Starykh2006, Chernyshev2009, Mourigal2013, Ma2016, Verresen2019, Macdougal2020,syromyatnikov2022} or on the interplay between magnons and fractionalized spinons~\cite{Mezio2011, Ghioldi2015, Ghioldi2018, Zhang2019, Ferrari2019, Zhang2020}. These theories offered a qualitative explanation to the downward renormalization of the three magnon bands. However, a comprehensive understanding to the dynamical spectra in the whole energy range, especially those in the intermediate- and high-energy scales, is still not available. In particular, it is unknown how the spectral weights are transferred to or from low-energy magnon excitations, damped longitudinal Higgs modes and high-energy continuum.
In this Letter, we resolve these problems through a thorough investigation on the spin-1/2 triangular Heisenberg model using a state-of-the-art tensor-network renormalization group method~\cite{Verstraete2004, Orus2019} in combination with the technique of automatic differentiation~\cite{Liao2019, Ponsioen2022}.

\bigskip
\textit{Model and method.---} 
\noindent
Ba$_3$CoSb$_2$O$_9$ has a highly symmetric hexagonal structure $P6_3/mmc$~\cite{Doi_2004}, and its magnetic Co$^{2+}$ ions with pseudospin $1/2$ form a perfect triangular lattice in the $ab$ plane.
It was proposed that this material presents an ideal realization of the paradigmatic spin-$1/2$ antiferromagnetic Heisenberg model in two dimensions~\cite{Susuki2013,Yamamoto2015,Shirata2012,Koutroulakis2015},
\begin{equation}
H = J \sum_{\langle i j\rangle}\left(S_{i}^{x} S_{j}^{x}+S_{i}^{y} S_{j}^{y}+\Delta S_{i}^{z} S_{j}^{z}\right),
\label{Ham}
\end{equation}
where $\langle ij \rangle$ runs over all the nearest-neighbor sites of the triangular lattice, $J$ is the antiferromagnetic coupling constant, and $\Delta$ is a parameter measuring anisotropy. In our calculation, we adopt the parameters determined from the magnetization, electronic spin resonance, nuclear magnetic resonance and neutron scattering measurements~\cite{Shirata2012, Susuki2013, Yamamoto2015, Koutroulakis2015, Ito2017, Kamiya2018, Macdougal2020}, namely,  $J=1.67$ meV and $\Delta=0.95$. We ignore the interlayer coupling because it is much smaller than the intralayer coupling~\cite{Susuki2013} and the observed  magnetic excitations are almost dispersionless along the $c$ axis~\cite{Ito2017}.

We employ the tensor-network formalism to simulate the magnetic excitation spectra
of Ba$_3$CoSb$_2$O$_9$ under the single-mode approximation~\cite{Feynman1954} in the framework of projected entangled pair states (PEPS). This approximation was first introduced in the framework of matrix product states by Ostlund and Rommer in one dimension~\cite{Ostlund1995, Haegeman2012}. It was extended to the PEPS presentation in two dimensions by Vanderstraeten \textit{et al}. ~\cite{Vanderstraeten2015}. Variational optimizations of local tensors are implemented with the approach of automatic differentiation first introduced to the tensor-network calculations in Ref.~\cite{Liao2019}. Recently, this approach was
extended to the calculation of excitation states in the single-mode approximation of PEPS~\cite{Ponsioen2022}.


The tensor-network calculation is based on the idea of renormalization group. It does not suffer from the notorious minus-sign problem encountered in the quantum Monte Carlo simulations and is applicable to a strongly correlated system with or without quantum or geometric frustrations, such as the model studied here. Moreover, it obeys the sum rule of spin fluctuation (see Fig.~S3 in Supplemental Material~\cite{SM} and Ref.~\cite{Ponsioen2022}) and can be directly applied to an infinite-lattice system without being bothered by the finite-size effect~\cite{SM}. 

\bigskip
\textit{Results.---} 
\noindent
Figure~\ref{fig1}(c) shows the intensity of the spin structural function
\begin{eqnarray}
&&  S(\boldsymbol k , \omega ) = \sum_\alpha S^{\alpha\alpha} (\boldsymbol k, \omega) , \quad (\alpha = x, y, z) \\
&&  S^{\alpha\beta} (\boldsymbol k, \omega) = \langle 0 | S_{-\boldsymbol k}^{\alpha} \delta(\omega-H+E_0)S_{\boldsymbol k}^{\beta} |0\rangle. 
\end{eqnarray}
calculated using the tensor-network methods along a representative path $\Gamma-K-M-K^{\prime}-\Gamma$ in the Brillouin zone. Here $S^{\alpha\alpha} (\boldsymbol k, \omega)$ is the dynamical spin-spin correlation function along the three axes. Three striking features are revealed in the spectra in different energy ranges.

\begin{figure}[tp]
\centering
\includegraphics[width=0.35\textwidth]{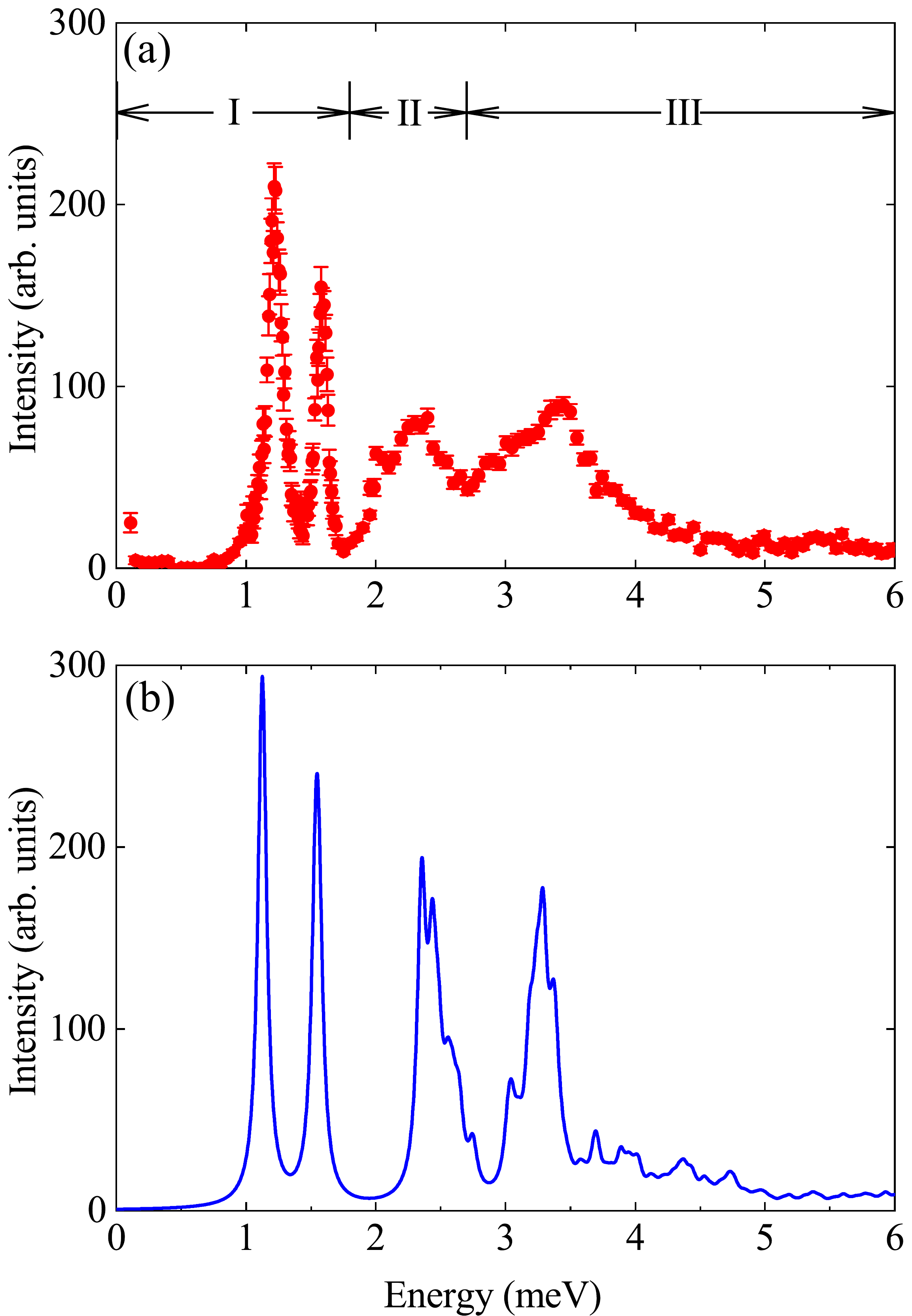}
\caption{
Comparison of the spectral weight at the $M$ point between the numerical calculation and the INS measurement.
(a) INS intensity for $\rm Ba_3CoSb_2O_9$, reproduced using the data published in Ref. \cite{Macdougal2020}.
(b) Numerical results obtained with the same parameters used for Fig.~\ref{fig1}.
\label{fig2}}
\end{figure}

\begin{figure*}[htp]
\centering
\includegraphics[width=0.85\textwidth]{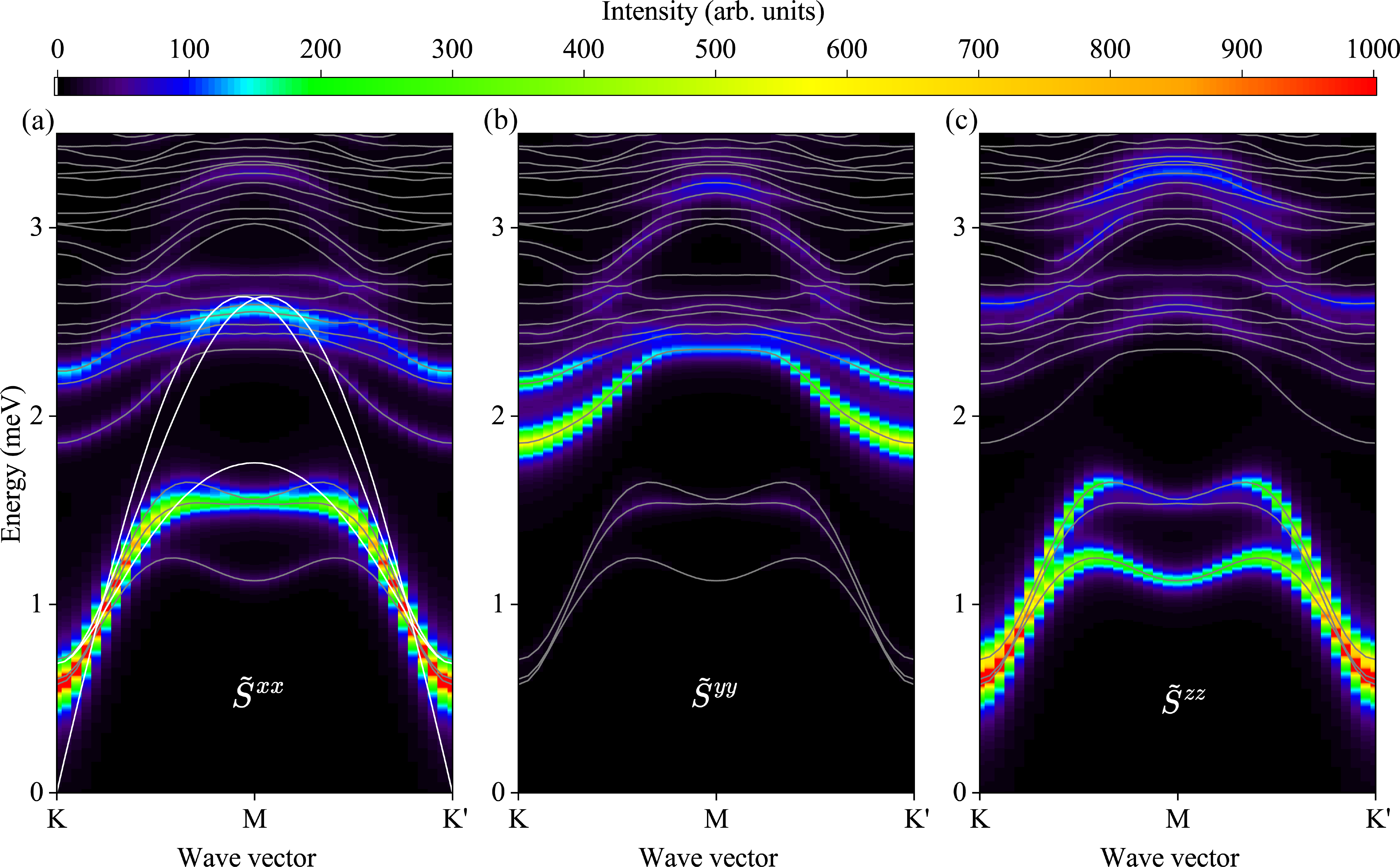}
\caption{
Sublattice spectral function in the magnetic Brillouin zone. The momentum runs along the primary vectors of the magnetic Brillouin zone that is equivalent to the path $K-M-K^{\prime}$ in the original Brillouin zone. The longitudinal direction is defined along the direction of the magnetization of sublattice $A$, i.e. along the $y$ axis. (a) In-plane transverse fluctuation modes $\tilde{S}^{xx}$. (b) Longitudinal fluctuation modes $\tilde{S}^{yy}$. (c) Out-of-plane transverse fluctuation modes $\tilde{S}^{zz}$. The gray curves show the energy dispersions of the excited states. 
The energy dispersions of magnons predicted by LSWT (white curves) are also shown in (a) for comparison.
\label{fig3}}
\end{figure*}

In the low-energy region, $\omega < 1.8$ meV, two sharp and one weak magnon excitation modes are observed (this can be seen more clearly from Fig.~\ref{fig3}). 
Around the $M$ point, only two sharp excitation modes are visible. The higher-energy mode is almost dispersionless, but the lower one exhibits a pronounced rotonlike minimum. By carefully examining the energy dispersions, shown by the gray lines in Fig.~\ref{fig1}(e), we find that these low-energy spectra are contributed by the three magnon bands, consistent with LSWT. However, the overall energy dispersions of these three magnon excitation modes deviate strongly from the LSWT prediction [see Fig.~S10 (e)]. The third excitation mode is not clearly seen because the spectral weight of the third band is very  small around $M$. This is consistent with the prediction of a resonating valence bond (RVB) theory~\cite{Zhang2020}. Around the $\Gamma$ point, the linear magnon dispersions are reproduced, but their intensities are very weak due to the cancellation inside a unit cell. Around the antiferromagnetic vector point $K$, the spectrum shows a sharp energy dispersion. This dispersion does not go to zero exactly at this point because the long-range correlation of the ground state is terminated by the finite virtual bond dimension of PEPS. Nevertheless, the lowest excitation energy gap at this point, as shown in Fig.~S9 in Supplemental Material \cite{SM}, tends to approach zero with the increase of the bond dimension, as a consequence of Goldstone's theorem. 
In the intermediate-energy region, $1.8$ meV$<\omega < 2.7$ meV, two $W$-like excitation modes are observed. These two modes are nearly energy degenerate at the $M$ point, rendering a strong coupling between these two modes. This energy range already falls within the two magnon excitation continuum. 
In the high energy region $\omega > 2.7$ meV, a weak and smeared $W$-like dispersive continuum whose tails extend to energy as high as 6 meV. The energy levels, shown in Fig.~\ref{fig1}(e), clearly become more densely packed in this energy range as an indication of excitation continuum. Many of them have invisible spectral weights. The intensity shows a relatively brighter spot at $\omega \sim 3.3$ meV around the $M$ point.

By comparison with the INS measurement data \cite{Ito2017,Macdougal2020}, shown in Fig.~\ref{fig1}(d), we find that the numerical result agrees very well with the experimental one in the whole energy range. This is a surprising result considering that there is not any adjustable parameter used in the calculation. In the low-energy region, the calculated dispersion relations of the three magnon bands agree quantitatively with the measurement data (see Fig.~S8 in Supplemental Material \cite{SM}). In the high-energy region, the INS spectrum looks more  diffusive than our numerical result. However, this does not mean that there is no feature in the INS data. In fact, in addition to the two sharp magnon peaks observed in the low-energy region, two more peaks are observed in the high-energy spectra of INS at the $M$ point \cite{Macdougal2020}. Figure \ref{fig2} compares theoretical results for the energy dependence of the intensity with the experimental one at that point~\cite{Macdougal2020}. Again, the four-peak spectrum with the peak energies obtained from the numerical calculation agrees well with the experimental ones. This is the first time the two broad spectral peaks above 2 meV are disclosed in a theoretical calculation. The two high-energy INS peaks are broader than the numerical ones. In Sec.～V in Supplemental Material~\cite{SM}, we present a detailed analysis of the bond-dimension dependence of the spin excitation spectra and find that these peaks are qualitatively unaltered and do not become significantly broadened with the increase of $D$. It suggests that the model Hamiltonian (\ref{Ham}) captures the main features of the spin excitation spectra of $\rm Ba_3CoSb_2O_9$ but is not sufficient to describe the diffusive INS spectra in the intermediate-energy scale. The derivation may result from the disorder, impurity, weak interlayer, or long-range interactions present in real materials but not included in the model Hamiltonian.  

The projected spectra functions along the three principal axes, $S^{xx}(\boldsymbol k,\omega)$, $S^{yy}(\boldsymbol k,\omega)$ and $S^{zz}(\boldsymbol k,\omega)$, can be measured by utilizing spin-polarized neutrons. This provides a unique approach to experimentally test our numerical predictions shown in Figs.~\ref{fig1}(e)-(g). As the magnetization is coplanar ordered, the low-energy spectral weight of $S^{zz}(\boldsymbol k,\omega)$ contributes mainly from the out-of-plane transverse fluctuations. The low-energy spectral weights of $S^{xx}(\boldsymbol k,\omega)$ and $S^{yy}(\boldsymbol k,\omega)$, on the other hand, contribute from the in-plane transverse as well as longitudinal fluctuations.

To further elucidate the microscopic nature of low-energy excitations, we investigate the sublattice spectral functions in the framework of coordinates where all magnetic ordered spins are locally rotated toward the positive direction of the $y$-axis. We denote the corresponding spectral functions as $\tilde S^{xx}(\boldsymbol k,\omega)$, $\tilde S^{yy}(\boldsymbol k,\omega)$ and $\tilde S^{zz}(\boldsymbol k,\omega)$. In this case, $\tilde S^{yy}(\boldsymbol k,\omega)$ contributes mainly from the longitudinal fluctuations, and $\tilde S^{xx}(\boldsymbol k,\omega)$ and $\tilde S^{zz}(\boldsymbol k,\omega)$ contribute mainly from the in-plane and our-of-plane transverse fluctuations, respectively.

Figure \ref{fig3} shows the numerical results for the three sublattice spectral functions $(\tilde S^{xx}, \tilde S^{yy}, \tilde S^{zz})$. Compared with the LSWT prediction (white curves in Fig. \ref{fig3}(a)), we find that the three magnon bands are strongly renormalized by their interaction with the high-energy continuum. For the three magnon bands below 1.8 meV, it is clear from Fig. \ref{fig3}(c) that the two bands whose energies are largest and smallest at the $M$ point result predominantly from the out-of-plane transverse spin fluctuations. The third magnon band, whose energy lies between the other two bands at the $M$ point, apparently results mainly from the in-plane spin fluctuations.

In the intermediate-energy region $1.8 < \omega < 2.7$ meV, Fig.~\ref{fig3}(b) clearly shows the two lowest-energy bands contribute mainly by the longitudinal spin fluctuations. This suggests that they are the damped Higgs modes, consistent with the prediction made based on the RVB picture~\cite{Zhang2020}. It can be verified experimentally by taking spin-polarized neutron scattering measurements.

Figure \ref{fig3} suggests that there is a significant transfer of spectral weights from the three low-energy magnon modes to the high-energy continuum. 
Such pronounced spectral weight transfer results inevitably from the interaction between the magnon excitations and the high-energy continuum. It leads to the downward bending of the three magnon bands around $M$, which implies that the interactions of these bands with other excitations around that point are very strong~\cite{Verresen2019, Ferrari2019, Zhang2020, Macdougal2020}. As the spectral weight of the highest-energy magnon band is almost completely suppressed around the $M$ point in all the directions, it further suggests that the downward bending of this band is not simply a consequence of the level repulsion imposed by high-energy excitation states. Otherwise, some remnant spectral weight from the original magnon band should be observed.

\bigskip
\textit{Discussion.---} 
\noindent
Our tensor network results reveal the key features of the dynamical spin spectra for the spin-1/2 antiferromagnetic Heisenberg model. Not only does it provide a good account for the INS spectrum of $\rm Ba_3CoSb_2O_9$, but also a comprehensive picture for understanding dynamical couplings between different excitation modes without invoking any approximation that is not easy to control. Our result of the sublattice spin structure factors shows unambiguously that the lowest-energy band whose intensity is more pronounced around the $K$ point in the intermediate-energy region contributes predominately by the longitudinal fluctuations, namely the Higgs modes. 
As the low-energy longitudinal fluctuations fall in the region of two-magnon continuum, their coupling with magnons leads to broadening of the Higgs peaks. The spectral peak around 2.3 meV in Fig.~\ref{fig2}(a) at the $M$ point, on the other hand, comes mainly from the in-plane transverse excitation mode, but strongly damped by its interaction with the Higgs mode. Besides, there are two kinds of transverse excitation modes, from the in-plane and out-of plane spin fluctuations, respectively. Their dynamical responses, as shown in Figs.~\ref{fig1}(e)-(g), can be differentiated by taking INS measurements with polarized neutrons. Furthermore, we find that the spin excitation spectra weakly depend on the anisotropic parameter $\Delta$ (see Fig.~S10 in Supplemental Material \cite{SM}).

This work demonstrates the great potential of the tensor network method in exploring the dynamic properties of highly frustrated antiferromagnets.
It offers a new tool to reveal the dynamic nature of exotic phases of quantum materials, such as quantum spin liquids and spin ices, and can be extended to study strongly correlated electronic systems~\cite{Ponsioen2022}. 
Further improvement to the numerical results can be done by increasing the bond dimension of local tensors. This can increase the accuracy of the PEPS wave functions, especially for the low-energy excitation modes with long correlation lengths, and improve the energy resolution of dynamical correlation functions.

We thank Tao Li, Chun Zhang, Yi-Bin Guo and Xuan Li for helpful discussions.
We thank Hidekazu Tanaka for providing the reference data.
This work is supported by the National Key Research and Development Project of China (Grant No.~2017YFA0302901), 
the National Natural Science Foundation of China (Grants No.~11888101, No.~11874095, and No.~11974396), 
the Youth Innovation Promotion Association CAS (Grant No.~2021004), and the Strategic 
Priority Research Program of Chinese Academy of Sciences (Grants No.~XDB33010100 and No.~XDB33020300).

\bibliographystyle{apsrev4-1}

\newpage

\clearpage

\addtolength{\oddsidemargin}{-0.75in}
\addtolength{\evensidemargin}{-0.75in}
\addtolength{\topmargin}{-0.725in}

\newcommand{\addpage}[1] {
 \begin{figure*}
   \includegraphics[width=8.5in,page=#1]{Appendix.pdf}
 \end{figure*}
}
\addpage{1}
\addpage{2}
\addpage{3}
\addpage{4}
\addpage{5}
\addpage{6}
\addpage{7}

\end{document}